\documentclass[5p,times]{elsarticle}
\usepackage{lipsum}
\usepackage{hyperref}
\journal{Physics Letters B}

\usepackage{cancel}
\usepackage{xspace}
\usepackage{caption}
\usepackage{subcaption}

\def\antibar#1{\ensuremath{#1\bar{#1}}}%

\def\ttbar{\antibar{t} \xspace }%
\newcommand{\madgraph}{\textsc{MG5\_aMC@NLO}}

\newcommand{\pythia}{\textsc{Pythia8}}

\newcommand{\delphes}{\textsc{Delphes}}

\newcommand{\vts}{\ensuremath{V_{ts}}}

\newcommand{\ks}{\ensuremath{K^0_S}}
\newcommand{\absvts}{\ensuremath{|\vts|}}

\newcommand{\ifb}{fb$^{-1}$}

\begin{document}

\begin{frontmatter}

\title{
Measuring $|V_{ts}|$ directly using strange-quark tagging at the LHC
}

\author{Woojin Jang}
\ead{wjang@physics.uos.ac.kr}
\author{Jason Sang Hun Lee}
\ead{jlee@physics.uos.ac.kr}
\author{Inkyu Park}
\ead{icpark@physics.uos.ac.kr}
\author{Ian James Watson}
\ead{ijwatson@physics.uos.ac.kr}
\address{University of Seoul,\\163, Seoulsiripdae-ro, Dongdaemun-gu, Seoul, Republic of Korea}

\begin{abstract}

The Cabibbo-Kobayashi-Maskawa (CKM) element \vts , representing the coupling between the top and strange quarks, is currently best determined through fits based on the unitarity of the CKM matrix, and measured indirectly through box-diagram oscillations, and loop-mediated rare decays of the $B$ or $K$ mesons.
It has been previously proposed to use the tree level decay of the $t$ quark to the $s$ quark to determine \absvts\ at the LHC, which has become a top factory.
In this paper, we extend the proposal by performing a detailed analysis of measuring $t \to sW$ in dileptonic \ttbar events.
In particular, we perform detector response simulation, including the reconstruction of \ks, which are used for tagging jets produced by $s$ quarks against the dominant $t \to bW$ decay.
We show that it should be possible to exclude $\absvts = 0$ at 6.0$\sigma$ with the expected High Luminosity LHC luminosity of 3000~\ifb.

\end{abstract}

\begin{keyword}
CKM matrix, s-jet tagging
\end{keyword}

\end{frontmatter}


\section{Introduction}

The Cabibbo-Kobayashi-Maskawa (CKM) matrix~\cite{Kobayashi:1973fv} gives the strength of the cross-generational weak couplings between the up and down type quarks, and is currently the only known source of charge-parity violation in the Standard Model (SM), which is required for understanding the observed matter--anti-matter asymmetry in the universe~\cite{Sakharov:1967dj}.
Measurements of the CKM matrix aim to overconstrain the matrix, testing the unitarity assumption.
Non-unitarity would indicate the existence of an additional, as yet unknown, coupling from Beyond the Standard Model (BSM) physics.
The CKM matrix element \vts determines the relative strength of the $t$ quark's weak decay to the $s$ quark compared to other down-type quarks.
The magnitude of \absvts\ determined through fits based on the unitarity of the CKM matrix is $39.78^{+0.82}_{-0.60} \times 10^{-3}$, and the indirect measurement through box diagram oscillations and rare decays involving loops is $38.8 \pm 1.1 \times 10^{-3}$~\cite{PDG}.
However, a recent reanalysis of Tevatron and 8~TeV LHC data has shown that after relaxing the unitarity constraints, \absvts\ can be as large as 0.1~\cite{Clerbaux:2018vup}.  
Additionally, a recent CMS analysis of 13~TeV data with the single top channel has given the constraint $\absvts + |V_{td}| < 0.057$ at the 95\%~CL under the SM assumption of CKM unitarity, and $\absvts + |V_{td}|=0.06 \pm 0.06$ after relaxing the unitarity constraints and allowing BSM contributions to the top width~\cite{Sirunyan:2020xoq}.
Further measurements are therefore required to constrain \absvts\ in the most general scenario, and in particular, the decay of $t \to sW$ has not yet been observed.

A direct measurement of $\absvts^2 = \frac{\mathcal{B} (t \to sW)}{\mathcal{B}(t \to qW)}$ at the LHC using the properties of strange hadrons to tag $s$ jets was proposed in~\cite{Ali_2010}.
That proposal made a generator level analysis to argue that the measurement would be feasible at the LHC, showing that under the assumption of perfect non-\ttbar background rejection and perfect top and hadron reconstruction, 10~\ifb~is sufficient to observe the $t \to sW$ decay.
Now that the LHC has collected more than an order of magnitude more luminosity than the proposal considered, we extend that study.
In particular, we perform a full reconstruction analysis using the \delphes\ fast detector simulation package, to make a more realistic estimate of the data luminosity required to observe the decay, and analyze the difficulties that would arise in such a measurement.
With our more realistic simulation setup, we investigate the prospects of measuring \absvts\ in several scenarios, including the future High Luminosity LHC (HL-LHC)~\cite{BejarAlonso:2020kmn}.

\section{Simulation and Event Selection}

We used \madgraph\ 2.4.2 to generate \ttbar events in the dilepton decay channel with up to 2 additional jets at next to leading order~\cite{Alwall_2014}.
We use the next-to-next-to-leading order top pair production cross-section $\sigma(pp \to t\bar{t})= 831.76$~pb for a collision energy of 13~TeV, which was calculated using the {\sc Top}++ program~\cite{Czakon:2011xx}.
We generated about 7 million signal \ttbar events where one of the $t$ quarks is forced to decay to a $s$ quark and about 7 million background \ttbar events are generated where both $t$ quarks decay to $b$ quarks.
Drell-Yan events with 2 additional partons are the dominant non-\ttbar backgrounds for dilepton \ttbar events.
We generated 20 million Drell-Yan plus two parton events for each of the 4 jet flavour categories: $bb$, $cc$, $ss$ and $qq~(q = u,d,g)$ and use the leading order cross section reported by \madgraph\ for the categories : $bb = 42.9$~pb, $cc = 4.31$~pb, $ss = 4.37$~pb and $qq = 23.8$~pb. In addition to the non-\ttbar backgrounds, we generated 10 million \ttbar plus one $s$ quark and two $s$ quarks respectively with the leading order cross section of 9.41~pb and 1.57~pb.
\pythia~8.212 was used to simulate parton showering and hadronization~\cite{Sjostrand_2015,Sjostrand_2006} with the FxFx merging scheme~\cite{Frederix:2012ps}.
We modified the \ks\ decay in \pythia\ to allow the \ks\ to decay inside a fiducial volume of a cylinder centered at the proton collision point with a radius of 860~mm and a length of 4400~mm. This is equivalent to the region of the CMS tracking detector where the pions from the decay may still pass through three silicon detectors, and would allow us to reconstruct \ks\ using reconstructed tracks when it decays to a charged pion pair.
We used \delphes~3.4.2 to simulate the response of a CMS-like detector with particle flow (PF) outputs~\cite{de_Favereau_2014}.
For jet clustering, we use the anti-$k_{t}$ algorithm with jet radius $R = 0.4$ using FastJet 3.3.2~\cite{Cacciari_2012}.
We used the default CMS card included in \delphes\ but updated it to match the CMS setup used of Run 2.
The jet radius was decreased from 0.5 to 0.4. 
The $\Delta R$ cone used to calculate lepton isolation was reduced from 0.5 to 0.3 for electrons and 0.5 to 0.4 for muons.
The track transverse momentum resolution formula was updated using the function given in~\cite{CERN-PH-EP-2014-070}. 
The b-tagging efficiency was updated to closely match the Run 2 response of CMS~\cite{CMS-PAS-BTV-15-001} and a $b$-tagging based on track counting module was added.
Smearing of the track impact parameter in the transverse plane was also added to emulate a more realistic \ks\ reconstruction using the associated module and parameters provided with \delphes, in order to replicate the performance of the CMS tracker~\cite{CMS-TRK-11-001}.


In this study, we use dilepton events in order to remove the additional jet activity from the $W$ decay, which additionally suppresses the background contribution from other processes, especially the multi-jet QCD background. 
The \ttbar event selection criteria are based on the CMS measurement of the top pair cross-section with the dilepton channel~\cite{CMS-TOP-17-014}.
First, we select events with two isolated leptons, each of which has $p_{T} >$ 25(20)~GeV for a leading (sub-leading) lepton and pseudorapidity $|\eta| <$ 2.4 and an invariant mass of a lepton pair is more than 20~GeV.
Then, we veto $Z$ boson production by excluding events in the dilepton invariant mass range of $|M_{Z} - M_{ll}| <$ 15~GeV, where $M_{Z} = 91.1876$ GeV~\cite{PDG}.
We require events to have missing energy $\cancel{E}_{T} >$ 40 GeV and at least two reconstructed jets with $p_{T} >$ 30 GeV and $|\eta| <$ 2.4.

We define primary jets as reconstructed jets which are matched to a generator level quark $q$ from the $t \to qW$ decay by finding the highest $p_T$ jet within $\Delta R < 0.4$ of the quark.
97\% of top pair production events have one primary jet, while 74\% of events have two primary jets, and therefore fully match the dilepton decay topology after reconstruction.
These primary jets will be used to train two Boosted Decision Trees (BDT).
The Toolkit for Multivariate Data Analysis in ROOT (TMVA) is used to train the BDT using the adaptive boosting algorithm~\cite{Hoecker_2007}.
The first BDT is trained to select the two primary jets out of all the reconstructed jets in the events.
The second BDT is trained to discriminate between $s$-quark initiated jets from other jets.
Once the first BDT selects the two primary jets, the second BDT is applied on these primary jets to look for $t \to sW$ decay.
This process is described in further detail below.
Both BDTs are trained using the signal $t\bar{t} \to sWbW$ and background $t\bar{t} \to bWbW$ samples.

Additional jets in \ttbar events create ambiguities in the selection of the primary jets, so we employ the first BDT to improve the efficiency of the primary jet selection.
We constructed a BDT model with the inputs of two jet and two lepton four vectors, the missing transverse momentum, and the $\Delta R$ between the two jets.
Each jet pair in the event is evaluated by the BDT.
We use the signal \ttbar sample to train this BDT, defining the signal to be the jet pair made from the two primary jets and the background is when one or more of the jets are not the primary jet.
The output of the first BDT is shown in Figure~\ref{fig:BDT1A}.
For each event, the jet pair with the highest BDT output is selected as the \ttbar primary jet candidates. Figure~\ref{fig:BDT1C} shows the top jet pair selection output for the signal, \ttbar, and Drell-Yan backgrounds. 

\begin{figure}[tb!]
     \centering
     \begin{subfigure}[b]{0.5\textwidth}
         \centering
         \includegraphics[width=\textwidth]{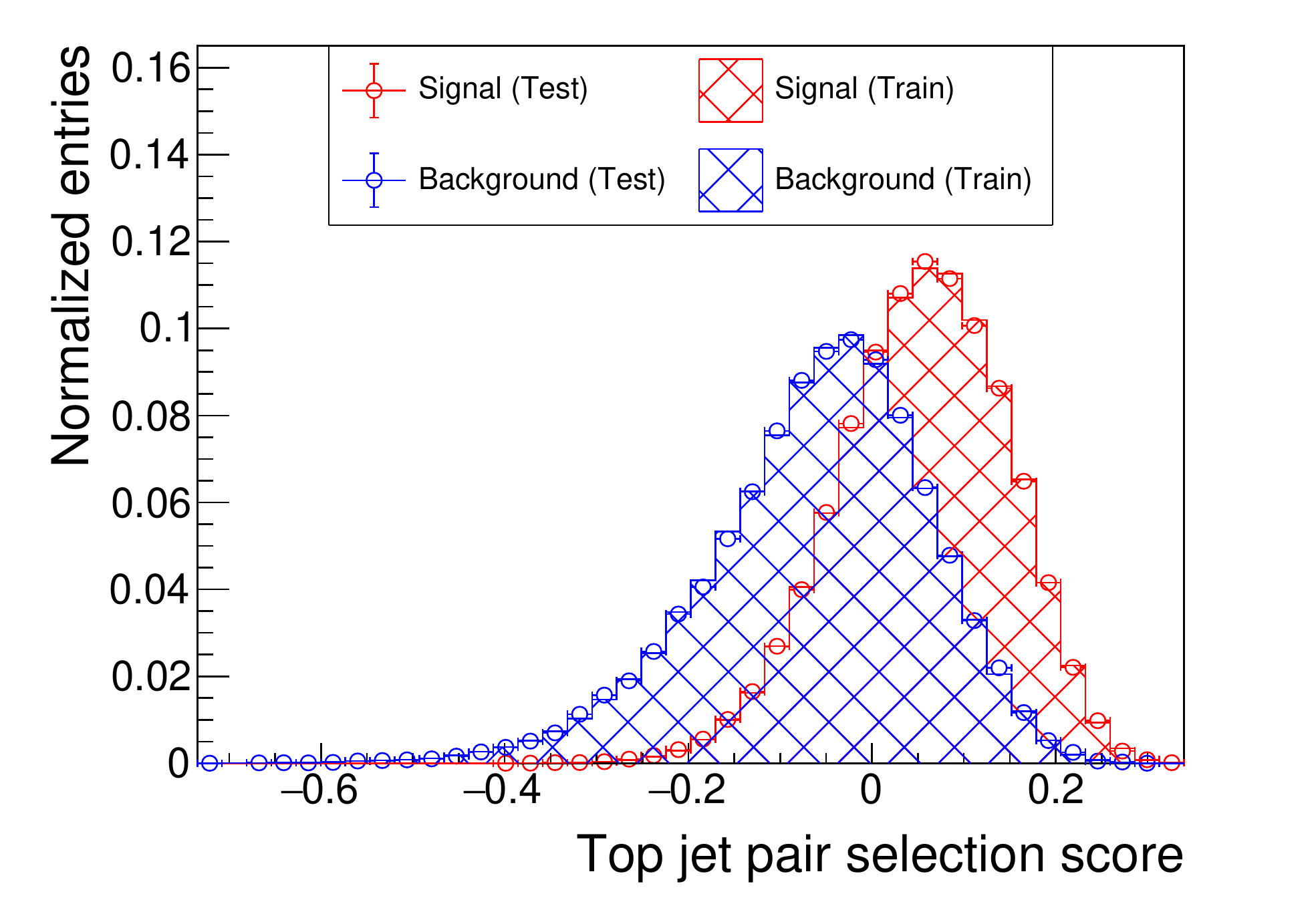}
         \caption{BDT output of primary jet selection}
         \label{fig:BDT1A}
     \end{subfigure}
     \hfill
     \begin{subfigure}[b]{0.5\textwidth}
         \centering
         \includegraphics[width=\textwidth]{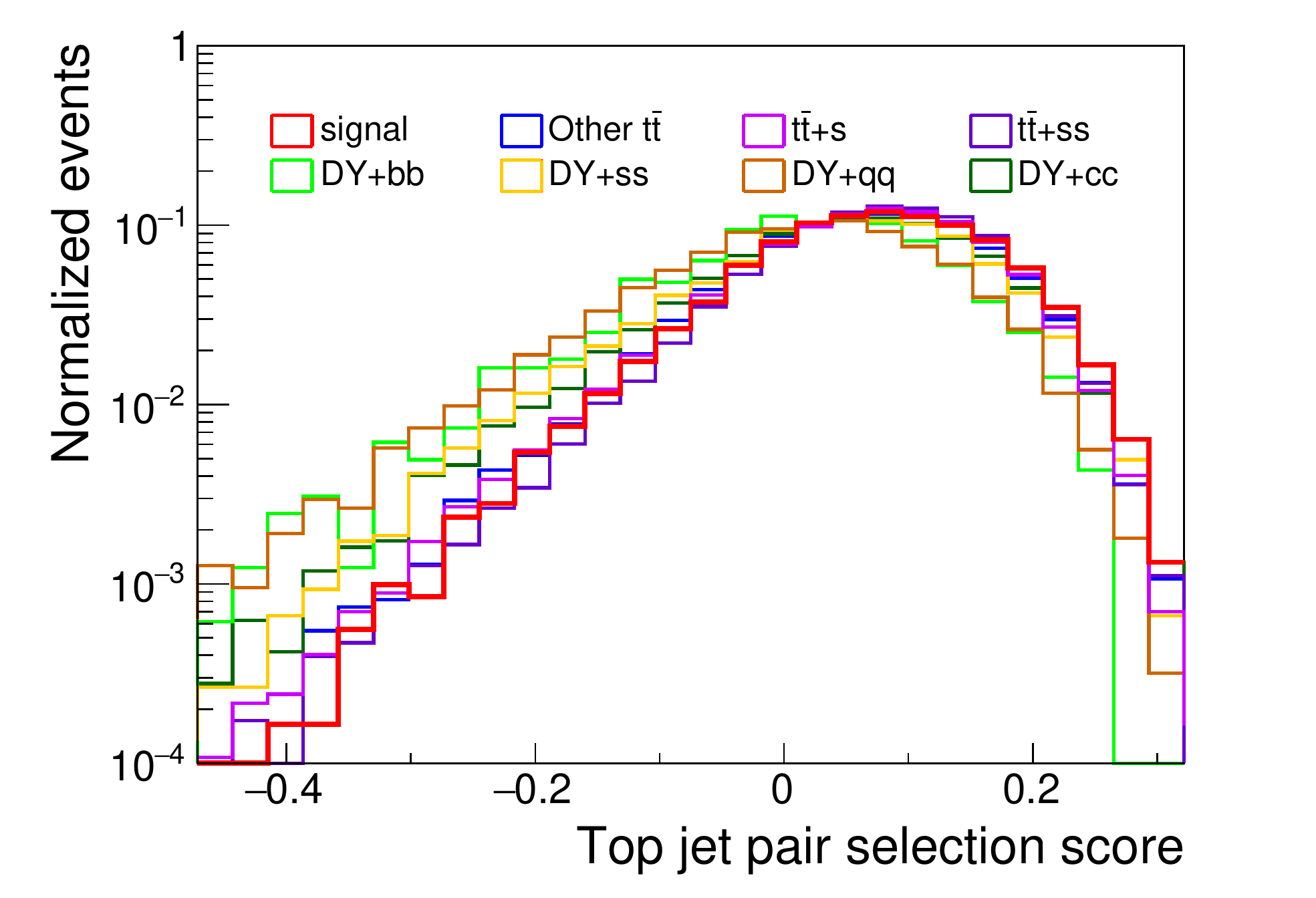}
         \caption{Highest BDT output of primary jet selection as density}
         \label{fig:BDT1B}
     \end{subfigure}
     \hfill
     \begin{subfigure}[b]{0.5\textwidth}
         \centering
         \includegraphics[width=\textwidth]{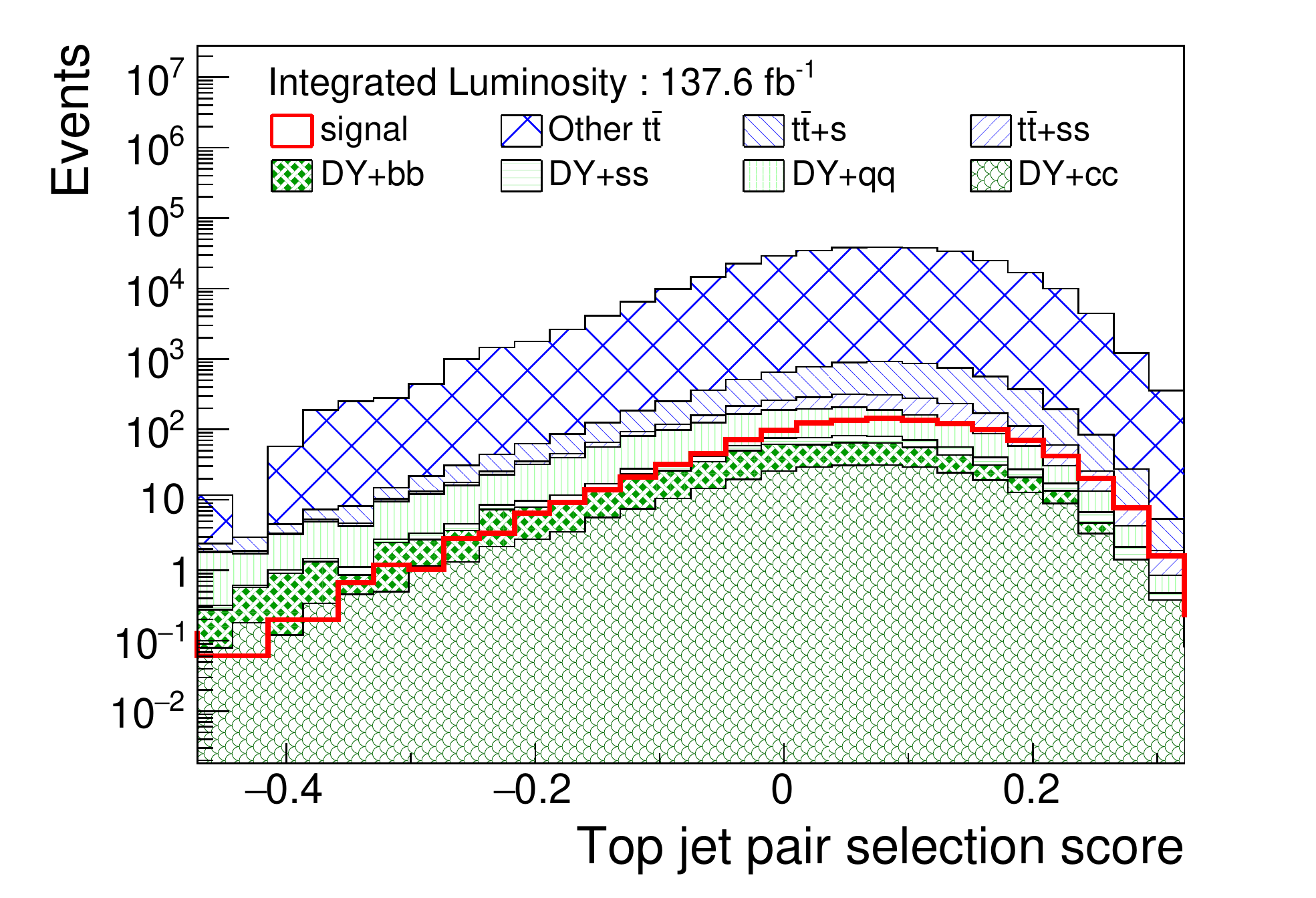}
         \caption{Highest BDT output of primary jet selection as expected number of event on the Run2 integrated luminosity}
         \label{fig:BDT1C}
     \end{subfigure}
        \caption{\ref{fig:BDT1A}) BDT output distribution of signal (Blue) and background (Red) on the BDT for the primary jet pair selection. \ref{fig:BDT1B}, \ref{fig:BDT1C}) BDT output distribution of jet pair selection for signal and background events normalized (\ref{fig:BDT1B}) and scaled to an integrated luminosity of 137.6 \ifb ~(\ref{fig:BDT1C})}
        \label{fig:top BDT three graphs}
\end{figure}

Next, in order to distinguish $s$ jets from the background, predominantly $b$ jets from the dominant $t\bar{t} \to bWbW$, we reconstruct \ks\ candidates inside the primary jet candidates.
In $s$ jets, \ks\ can be produced directly from the initiating $s$ quark, whereas in $b$ jets they will be produced after a cascade of decays of the $b$ hadron or from the quarks produced in the parton shower.
This means that \ks\ should be harder (relative to the jet energy) and more collimated in the case of $s$ quark initiated jets.
We reconstruct \ks\ using its decay into oppositely charged pion pairs, and due to the long lifetime of the \ks, we require the tracks to come from a displaced vertex within the tracker volume.
Using the charged hadron objects from the \delphes\ particle flow reconstruction, we consider all oppositely charged hadron pairs. 
Since general purpose detectors like CMS do not distinguish between pions from other charged hadrons, we assume all charged hadrons to be pions.
We require the charged hadron pair to have $p_{T} >$ 0.95~GeV and $|\eta| <$ 2.4 and the significance of the transverse impact parameter of each track to be greater than two, to ensure the tracks are not from the primary vertex.
Then we select reconstructed \ks\ candidates with an invariant mass of $|M_{\ks} - M_{\pi\pi}| < 0.1$~GeV, where $M_{\ks} = 497.611$~MeV~\cite{PDG}.
We check that the reconstructed \ks\ candidate is from a primary jet candidate by requiring the angle between the candidate momentum and the jet axis to satisfy $\Delta{}R < 0.4$.
If there are more than one reconstructed \ks\ candidates, we select the \ks\ with the highest $p_{T}$.
\begin{figure}[tb!]
     \centering
     \begin{subfigure}[b]{0.5\textwidth}
         \centering
         \includegraphics[width=\textwidth]{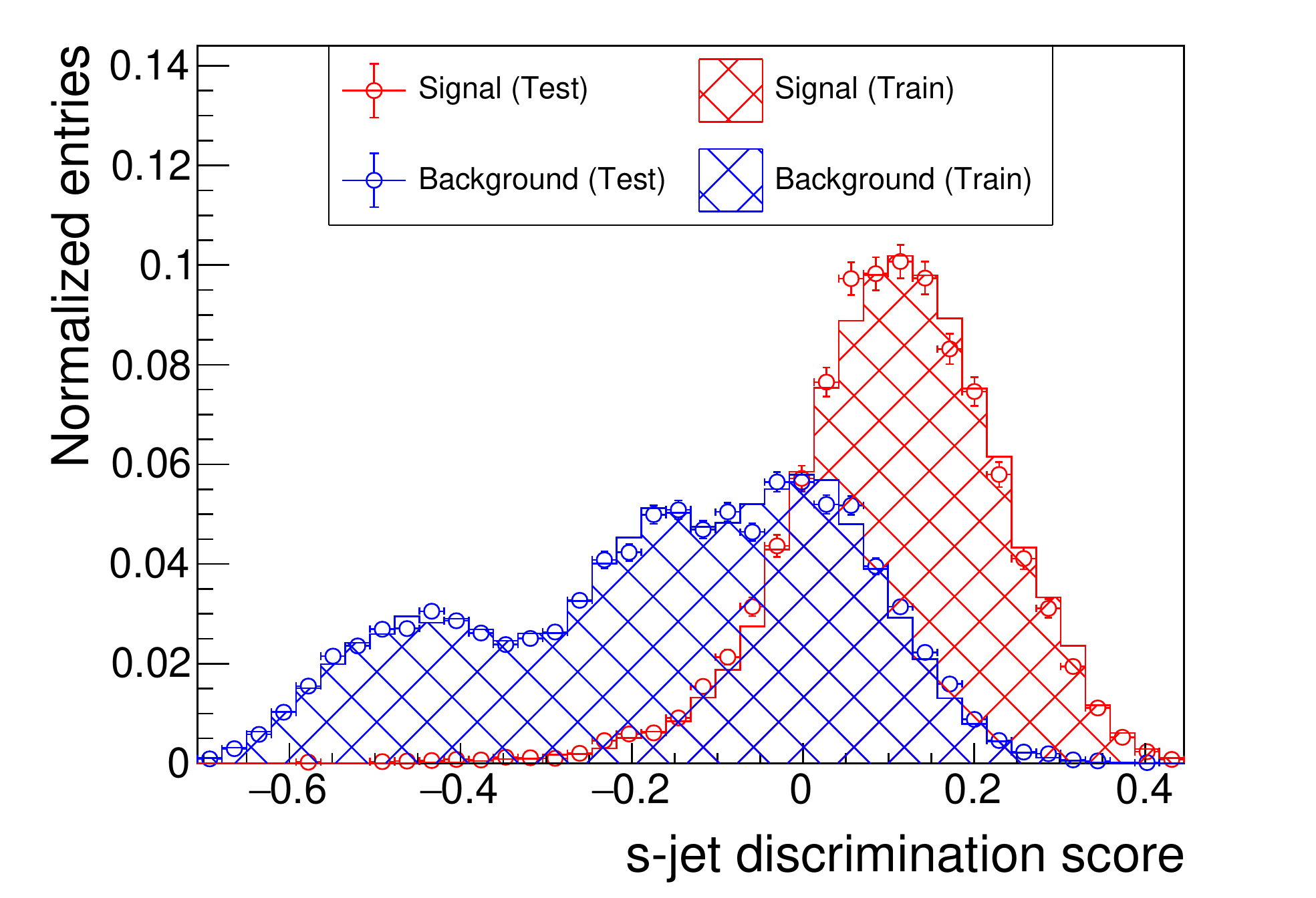}
         \caption{BDT output of $s$ jet tagging}
         \label{fig:BDT2A}
     \end{subfigure}
     \hfill
     \begin{subfigure}[b]{0.5\textwidth}
         \centering
         \includegraphics[width=\textwidth]{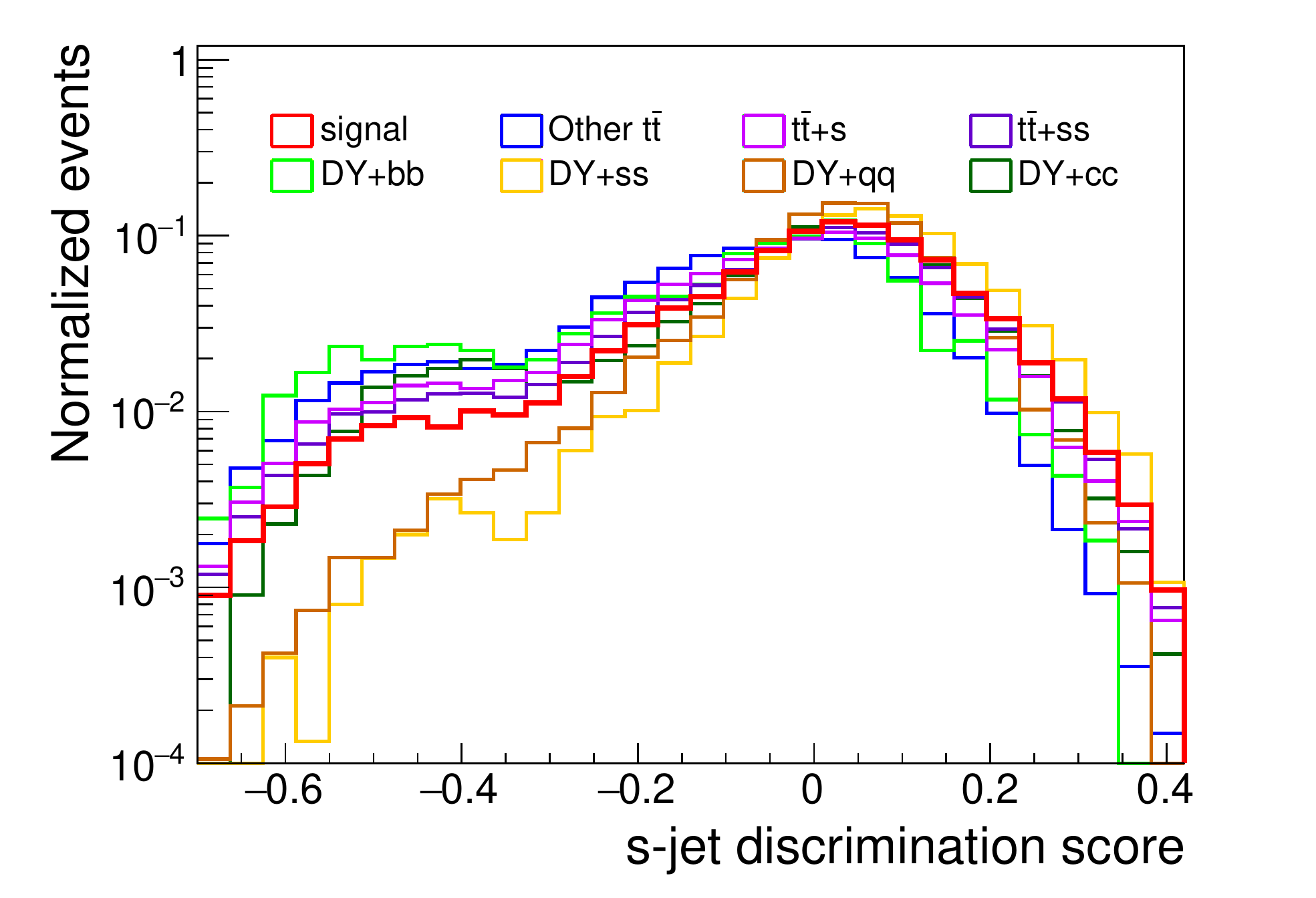}
         \caption{BDT output of $s$ jet tagging on signal and background events as density}
         \label{fig:BDT2B}
     \end{subfigure}
     \hfill
     \begin{subfigure}[b]{0.5\textwidth}
         \centering
         \includegraphics[width=\textwidth]{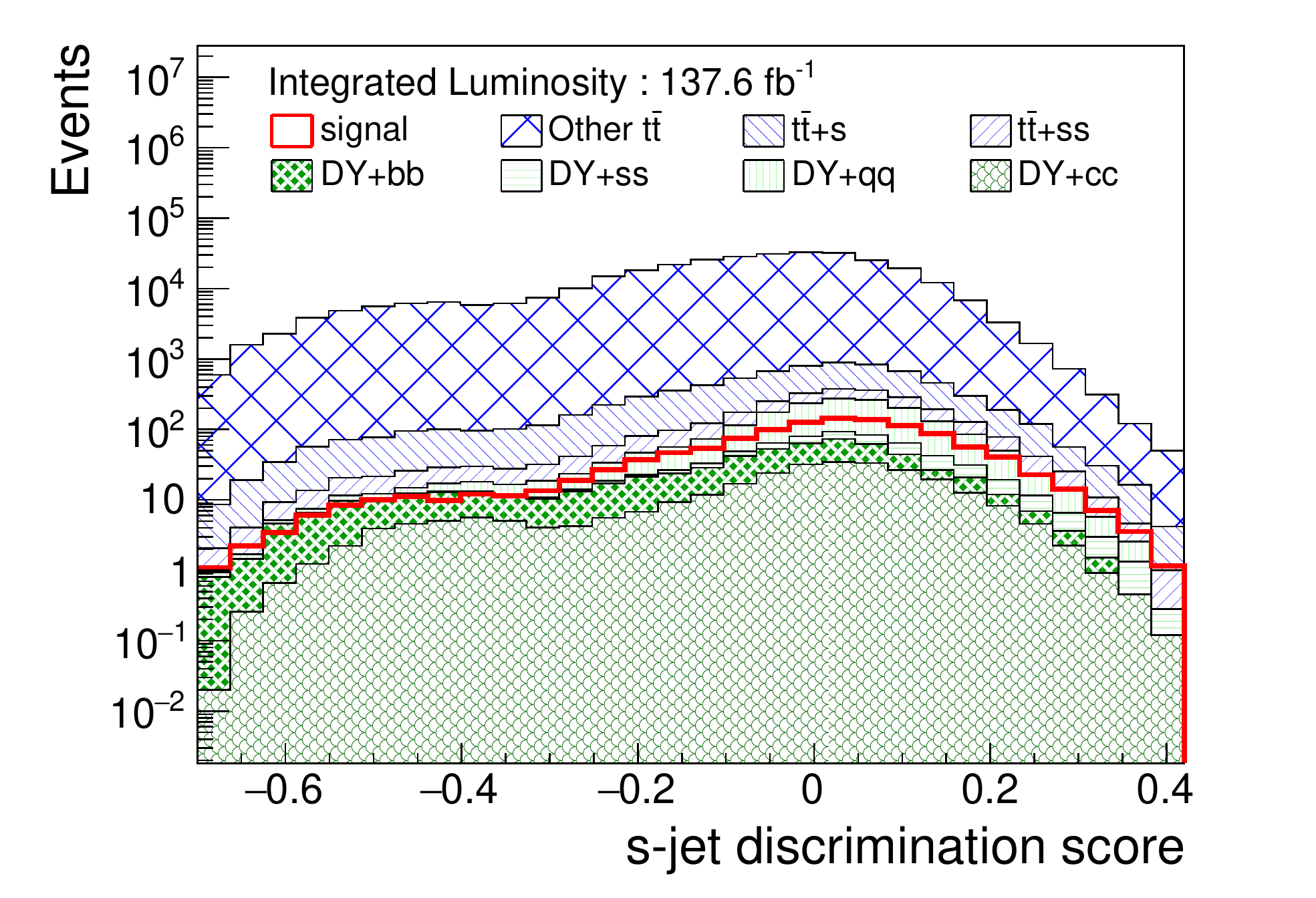}
         \caption{BDT output of $s$ jet tagging on signal and background events of the Run2 integrated luminosity}
         \label{fig:BDT2C}
     \end{subfigure}
        \caption{\ref{fig:BDT2A}) BDT output of $s$ jet tagging on signal (Blue) and background (Red). \ref{fig:BDT2B}, \ref{fig:BDT2C}) Highest BDT output of $s$ jet tagging on the primary jets normalized (\ref{fig:BDT2B}) and scaled to an integrated luminosity of 137.6 \ifb (\ref{fig:BDT2C})}
        \label{fig:jet BDT three graphs}
\end{figure}

After matching the \ks\ candidates to the primary jet candidates, we use both hadron and jet information to discriminate $s$ jets from all other jets.
First, from the jet information, we use jet's $p_{T}$, mass, its minor and major axes and their substructure-related quantities such as charged jet multiplicity, charged daughter's $p_{T}$ fraction in a jet, leptonic constituent's $p_{T}$ fraction in a jet, and $p_{T}D$, also called the jet energy sharing, defined as $\frac{\sqrt{\sum{p_{T,i}^2}}}{\sum{p_{T,i}}}$. 
To simulate b-tagging, we used a simple track counting method, which tags a jet as a $b$ if more than two tracks are found with a high impact parameter. 
From the \ks\ kinematics, we use the hadron's $p_{T}$ fraction relative to the jet $x = p_T^{had} / p_T^{jet}$, the $\Delta$R between the jet axis and the hadron momentum, the Distance of Closest Approach (DCA) between the two tracks, the cosine of the 2D pointing angle between momentum vector and position vector of hadron and decay length calculated by assuming the \ks\ vertex is the midpoint of DCA between the two charged tracks. In addition, the following information from both the charged pion daughters are included: the $p_{T}$ fraction compared to the \ks, the significance of the transverse impact parameter, and of the longitudinal impact parameter. 
For training the second BDT, the signal is defined as jets matched to an s-quark from the $t \to sW$ decay with a matching \ks\ whose momentum fraction, x $>$ 0.15 and the background is all other jets with \ks\ with x $>$ 0.15.
The result from the s-jet discriminating BDT training is shown in Figure~\ref{fig:BDT2A} and Figure~\ref{fig:BDT2C} shows the s-jet discriminating BDT for the primary jets on the signal and background processes. 

For the final $s$ jet selection, we first reject primary jets that are $b$ tagged in each event.
If both the primary jets are $b$ tagged, the event is vetoed from the analysis.
Of the remaining primary jets in each event, we select the jet with the highest $s$ tagging BDT output.
Figure~\ref{fig:BDT2C} shows the final $s$ tagging BDT output for signal and background events.

\section{Results}

The expected number of signal \ttbar and background events is given by
\begin{equation}
        \label{eq:Nexp:1}
        N_{sig} = \sigma(t\bar{t}) \times \mathcal{L} \times \mathcal{B}(t\bar{t} \to ql^{+}\bar{\nu}\bar{q}l^{-}\nu) \times 2|V_{ts}|^{2}|V_{tb}|^{2} \times \epsilon_{sig}
\end{equation}
\begin{equation}
        \label{eq:Nexp:2}
        N_{bkg} = \sigma(t\bar{t}) \times \mathcal{L} \times \mathcal{B}(t\bar{t} \to ql^{+}\bar{\nu}\bar{q}l^{-}\nu) \times |V_{tb}|^{4} \times \epsilon_{bkg} + N_{DY}
\end{equation}
where $\sigma(\ttbar)$ is the cross-section, $\mathcal{L}$ is a integrated luminosity, $\mathcal{B}(\ttbar \to ql^{+}\bar{\nu}\bar{q}l^{-}\nu)$ is the branching ratio of dileptonic decay mode, $\epsilon_{sig}$ ($\epsilon_{bkg}$) is the selection efficiency after the selections described in the previous section for the signal $sWbW$ (background $bWbW$) \ttbar sample, and $N_{DY}$ is the expected number of Drell-Yan background events.

\begin{figure}[th!]
\centering
    \includegraphics[width=.5\textwidth,trim=0 0 0 0,clip]{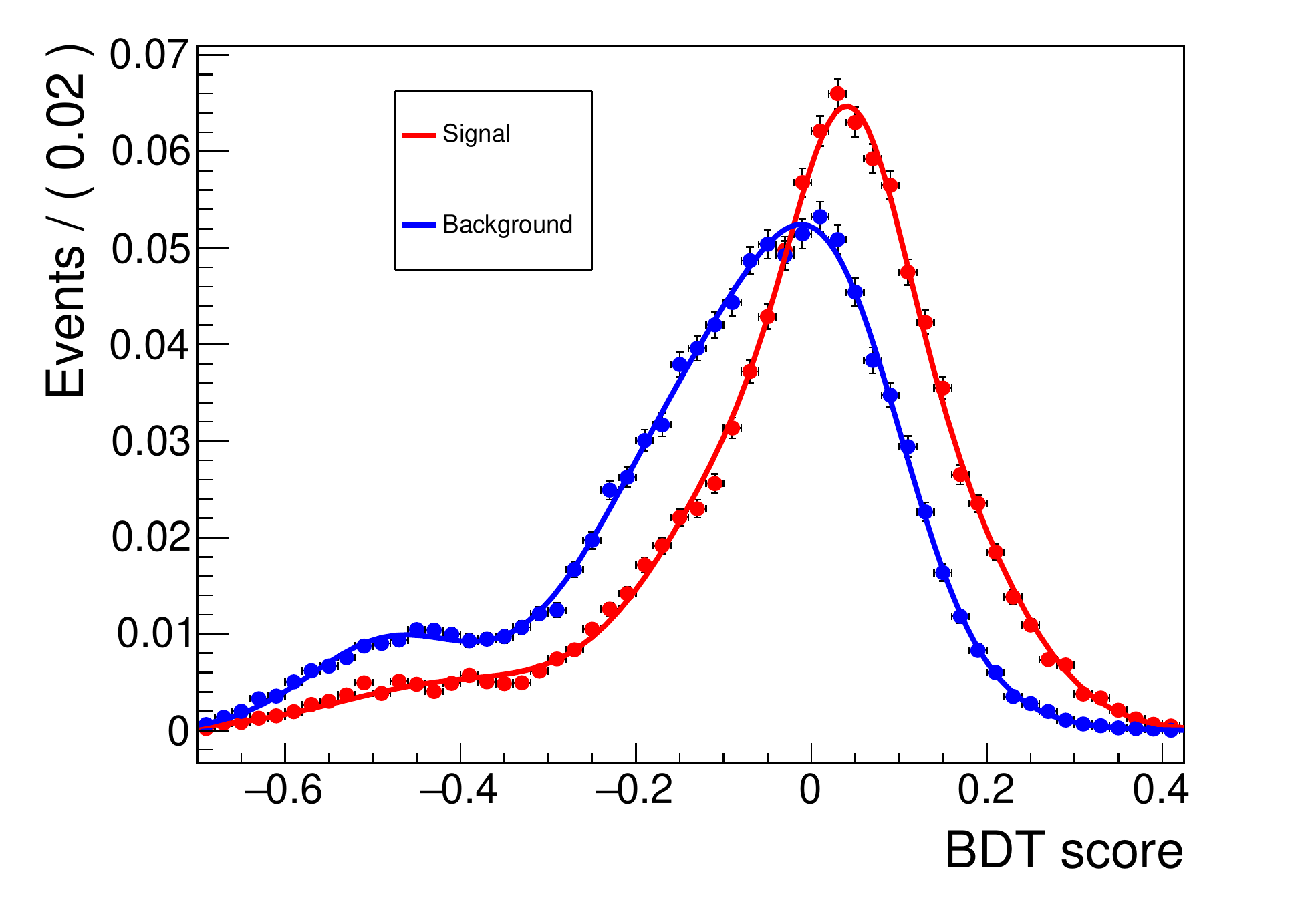}
    \caption{The gaussian fit of the BDT output distribution. The background distribution is shown in blue and the signal in red. The background includes Drell-Yan events with 2 jets as well as $t\bar{t} \to bWbW$ and the signal is $t\bar{t} \to sWbW$.
    }
    \label{fig:gaussianfit}
\end{figure}

\begin{figure*}[th!]
    \includegraphics[width=.5\textwidth,trim=0 0 0 0,clip]{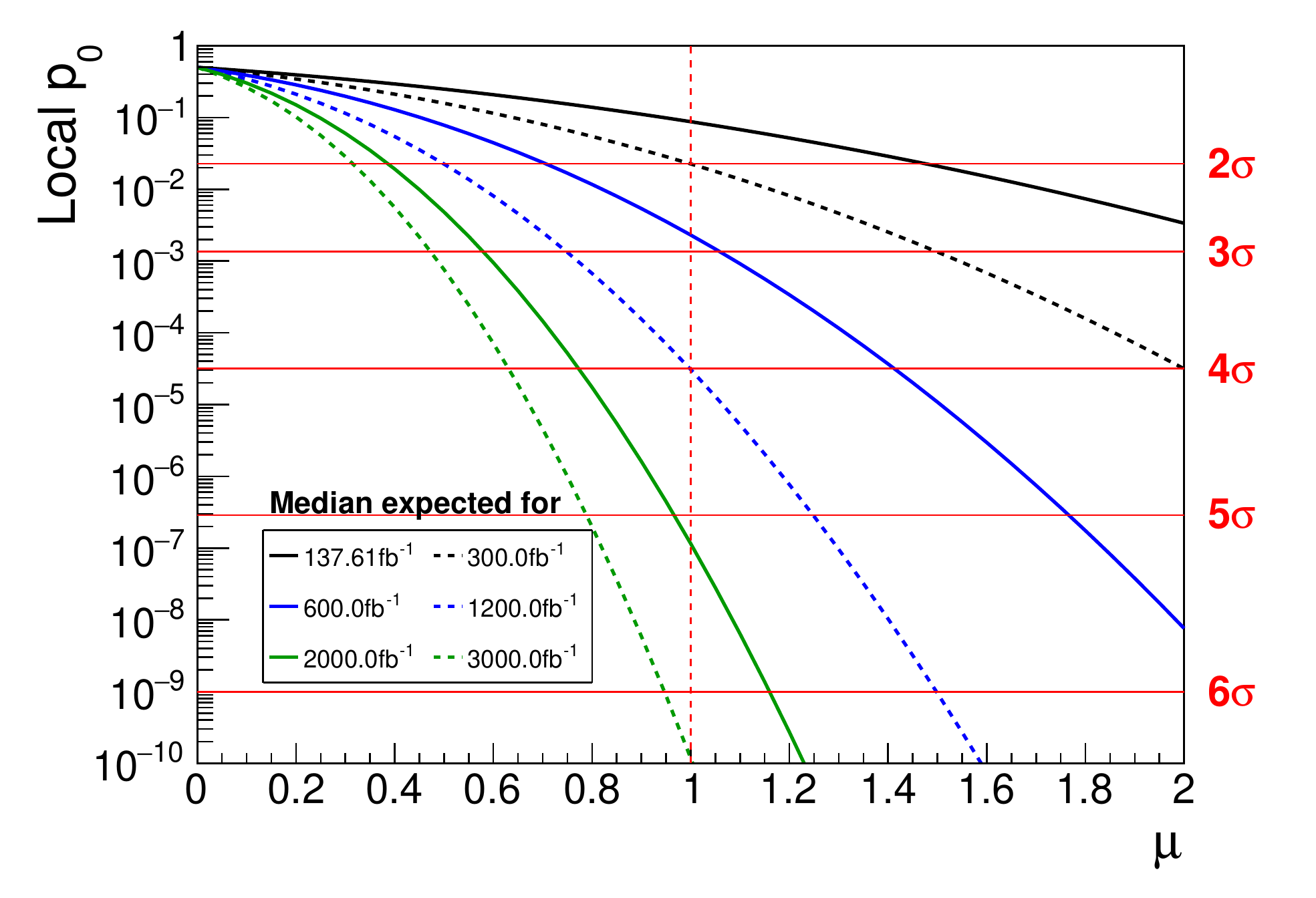}
    \includegraphics[width=.5\textwidth,trim=0 0 0 0,clip]{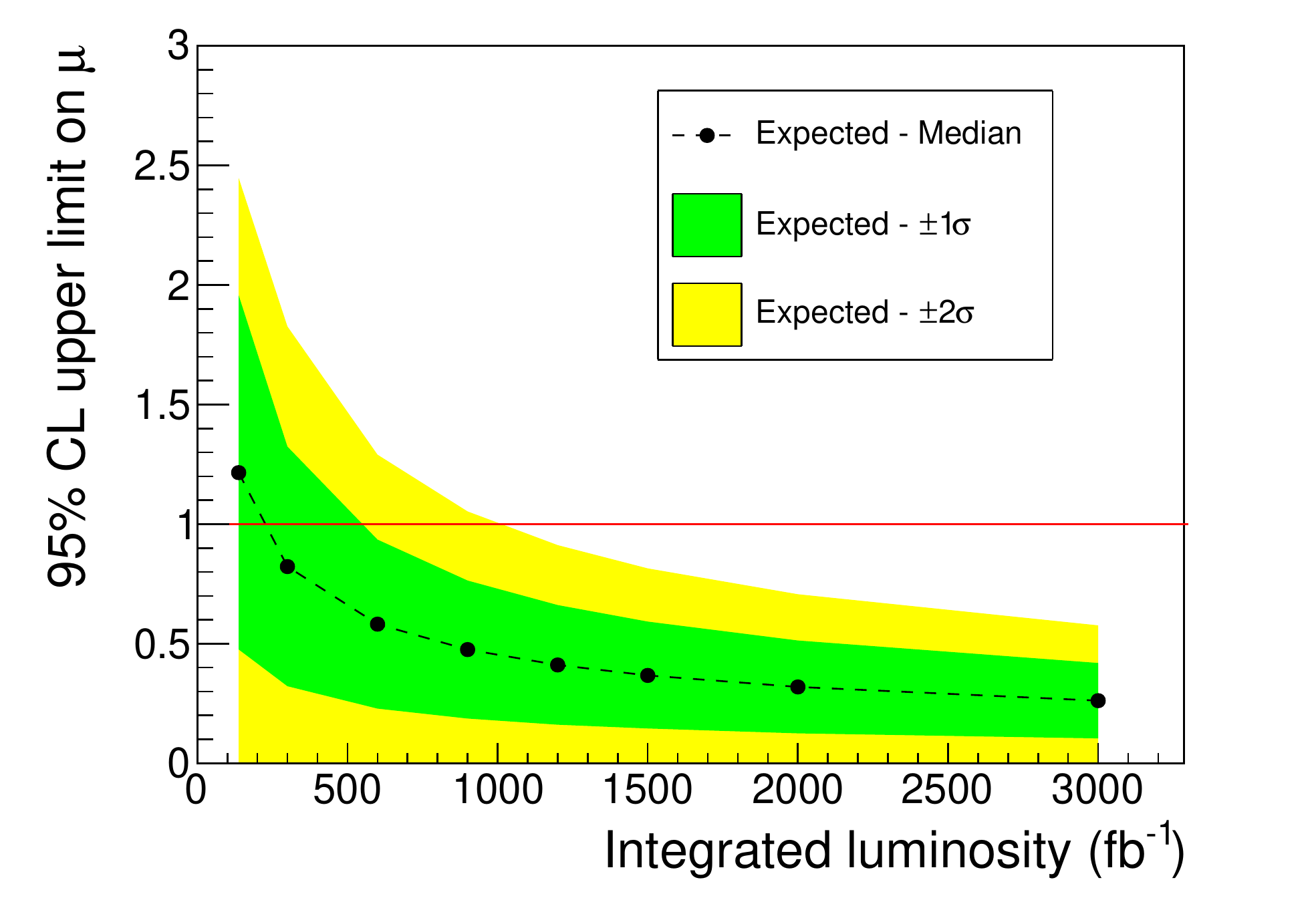}
    \caption{Local p$_{0}$ value (Left) and 95\% Confidence Level (CL) upper limit on signal strength $\mu$ (Right) for each integrated luminosity. In the left plot, p$_{0}$ is p value for background-only PDF and the red dashed line is on $\mu = 1$. Black dots in the right plot are $\mu$ corresponding to CL of 95\% at 137 (Run2), 300 (expected for Run3), 600, 1200, 2000 and 3000 (expected for HL-HLC) \ifb.  }
    \label{fig:Limit}
\end{figure*}

\begin{figure*}[th!]
    \includegraphics[width=.5\textwidth,trim=0 0 0 0,clip]{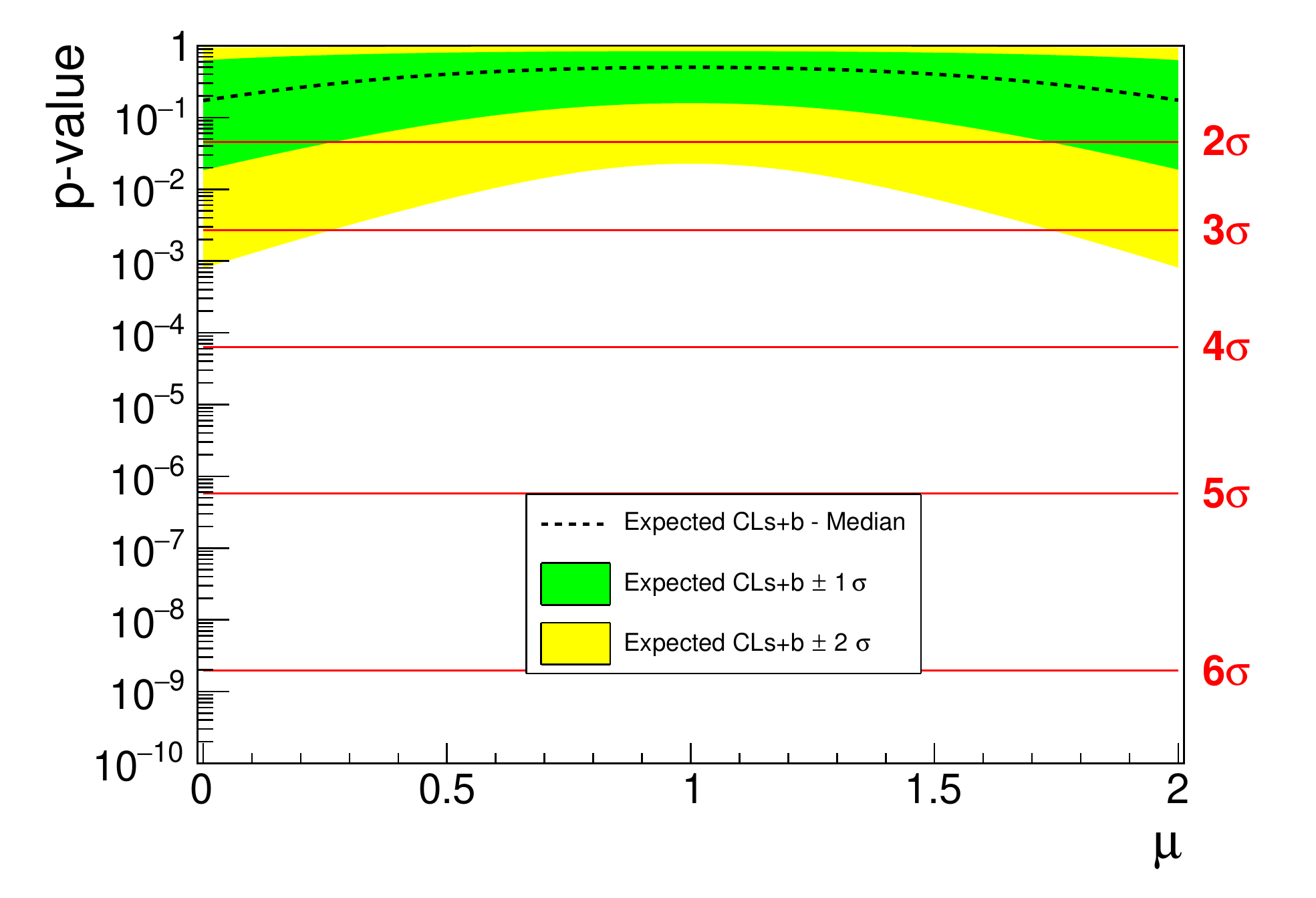}
    \includegraphics[width=.5\textwidth,trim=0 0 0 0,clip]{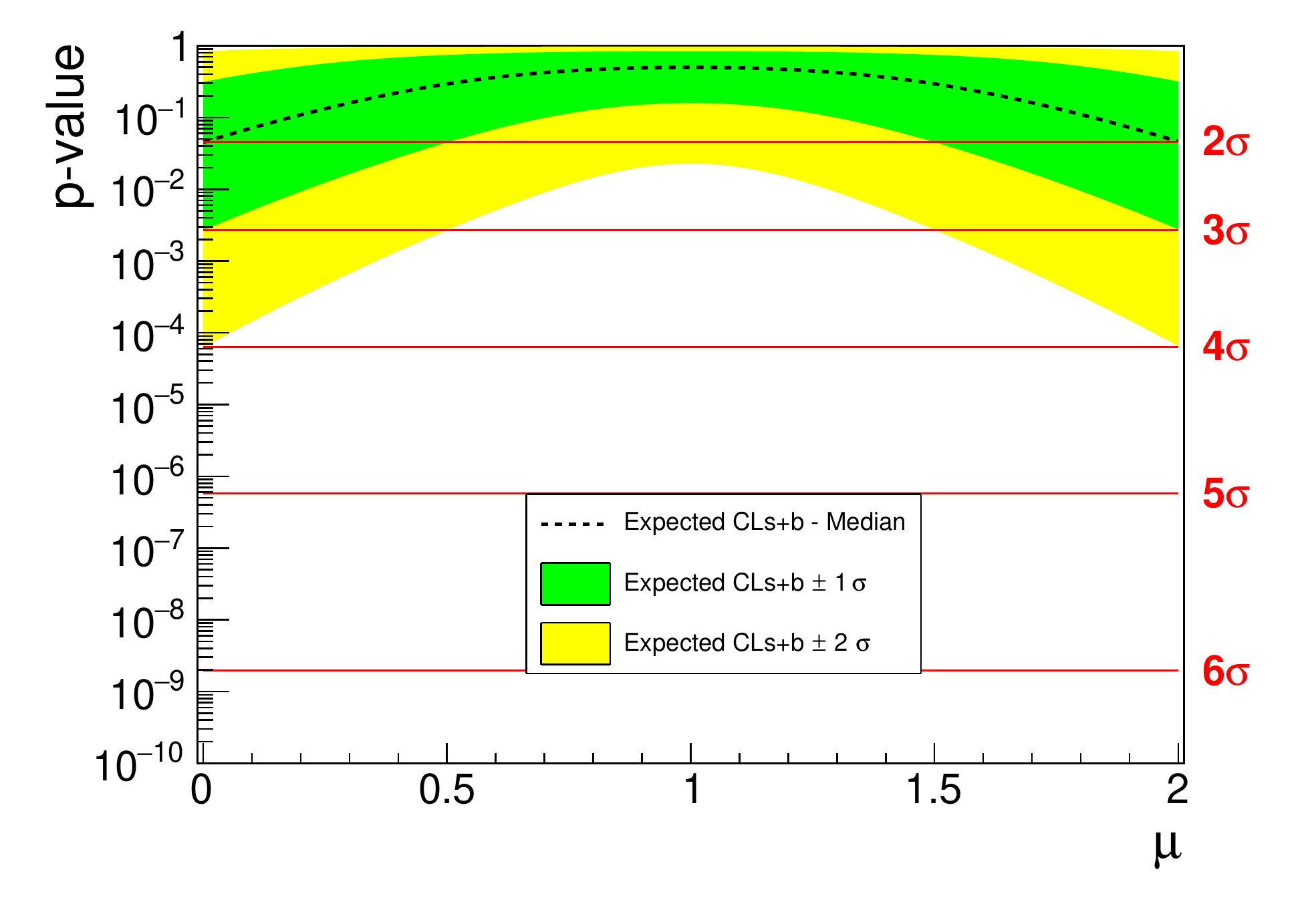}
    \includegraphics[width=.5\textwidth,trim=0 0 0 0,clip]{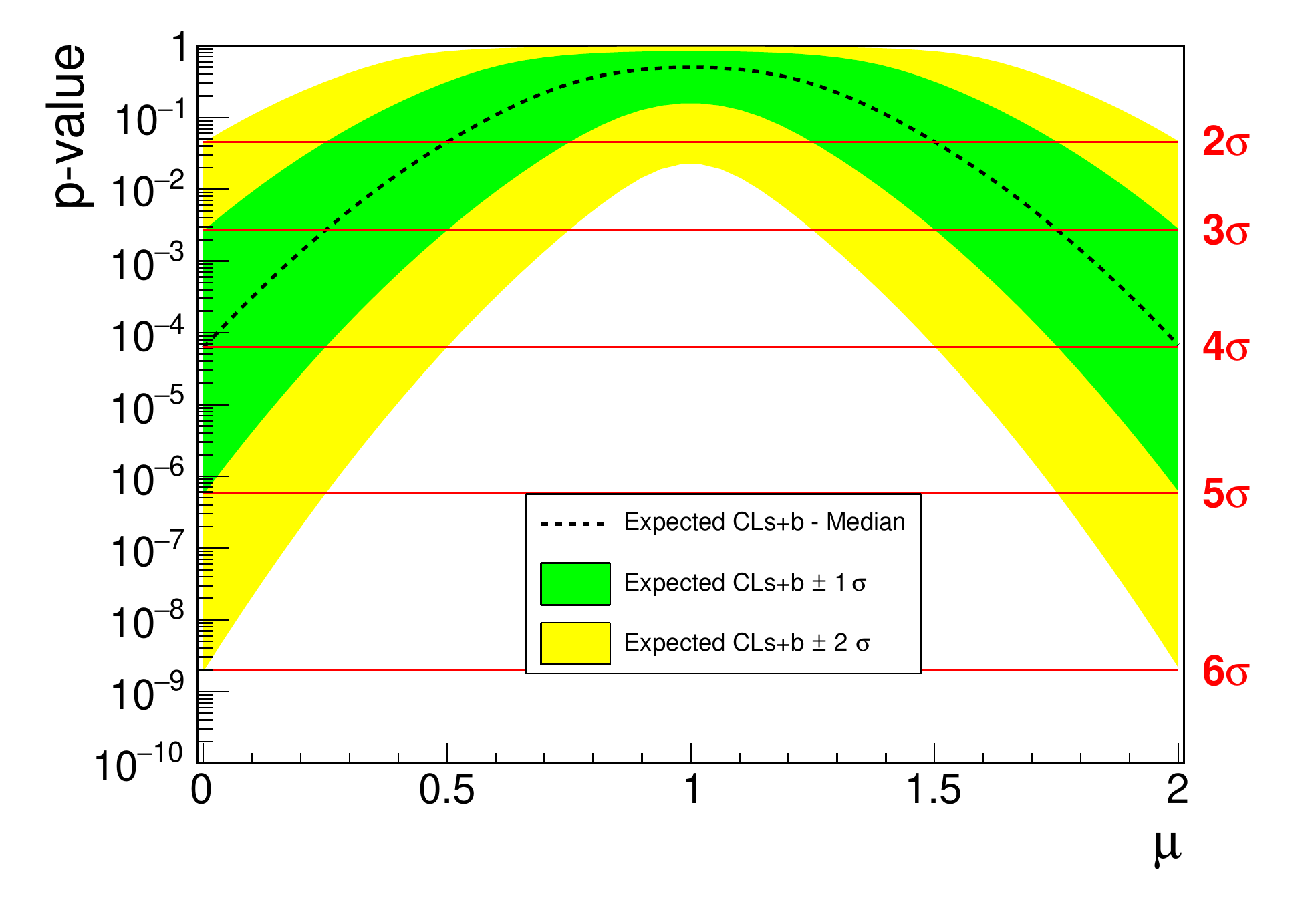}
    \includegraphics[width=.5\textwidth,trim=0 0 0 0,clip]{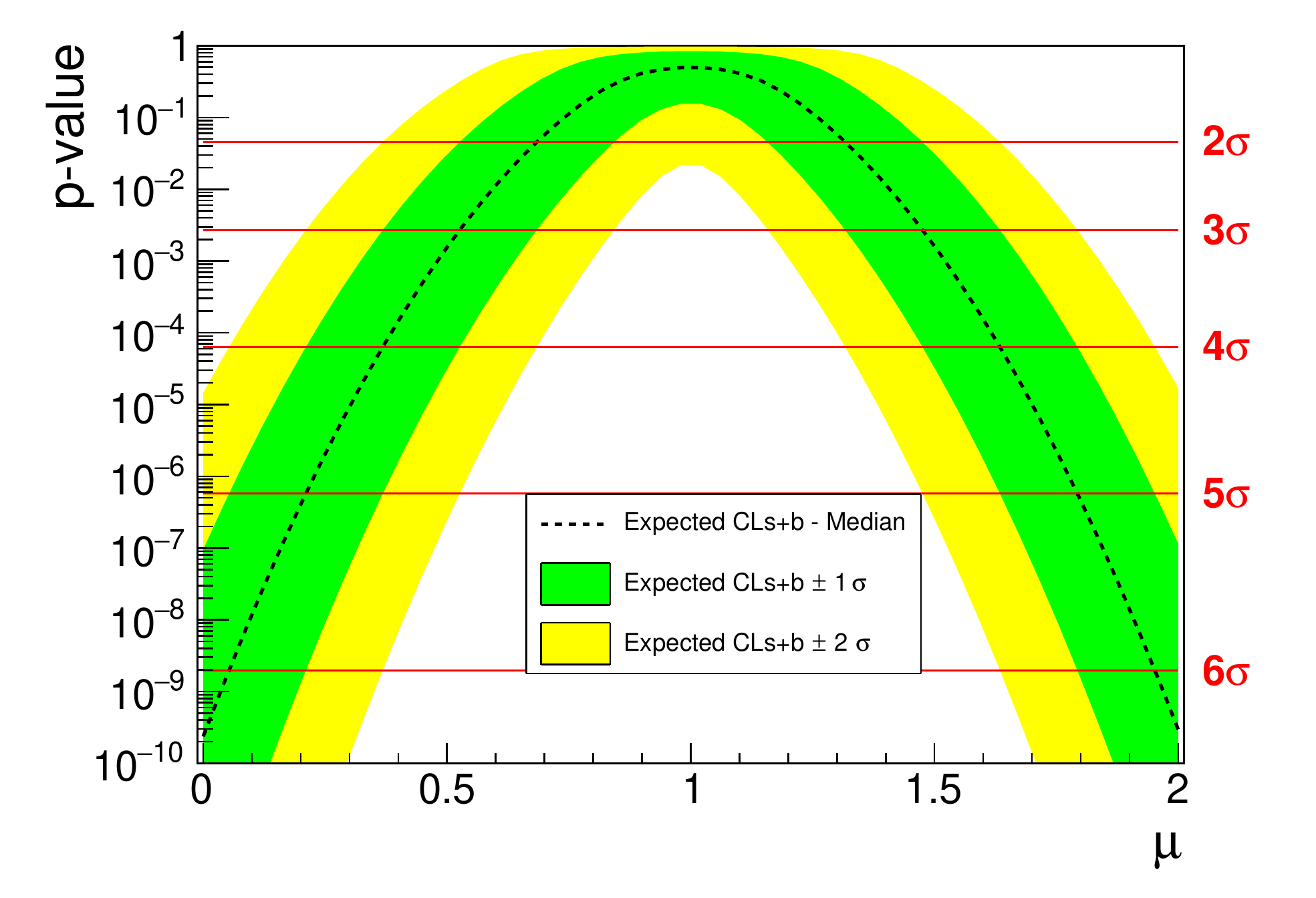}
    \caption{$p$-value at the Run2 luminosity (top left), the expected luminosity from Run3 (top right), 1200 fb$^{-1}$ (bottom left) and 3000~\ifb\ the design luminosity of HL-LHC (bottom right). The dotted line is the median of the expected CL$_{s+b}$, the green (yellow) band is  $\pm 1\sigma$ ($\pm 2\sigma$) range.}
    \label{fig:CLsb}
\end{figure*}

To study the feasibility of a direct measurement of $\absvts$, we perform an analysis to find the expected significance to reject the hypothesis $H_0$ of no $t \to sW$ decays, the expected upper limit on $\absvts$ and also an expected confidence interval $\absvts$.
We use \textsc{RooFit}~\cite{Verkerke_2003} and \textsc{RooStats}~\cite{Moneta_2010} to perform the statistical analysis using the \textsc{RooStats} asymptotic calculator based on the asymptotic properties of likelihood function~\cite{Cowan_2011}.
We fit the $s$ jet tagging BDT distributions from Figure~\ref{fig:BDT2C} to obtain approximations of the probability density functions (PDFs) for further study.
The PDFs for the signal and background are separately modeled by a sum of four Gaussian distributions and the fitted PDFs are shown in Figure~\ref{fig:gaussianfit}.
The BDT distribution in this figure has a different definition of signal and background from the BDT distribution of jet discriminator.
The BDT distribution shown in the previous section defines the signal (background) as jets matched to generator level s (b) quarks, while the distributions here define the signal as the jet with the highest BDT output in an event from $t\bar{t} \to sWbW$ signal events and the background as the jet with the highest output from $t\bar{t} \to bWbW$ and Drell-Yan events.
The total model is thus 
\begin{equation}
    \label{eq:PDF}
    P_{s+b} = \mu \times N_{sig} \times P_{sig} + \nu \times N_{bkg} \times P_{bkg}
\end{equation} 
where $N_{sig}$ is the number of $\ttbar \to sWbW$ events expected by the SM, corrected for the selection efficiency and $\mu$ is the signal strength relative to the SM $\frac{|V_{ts}^{obs}|^2}{|V_{ts}^{CKM}|^2}$, where $|V_{ts}^{CKM}| = 39.78 \times 10^{-3}$~\cite{PDG}, and $\nu$ is a nuisance parameter to control the background level relative to the expected background.
From the model PDF, we generate an Asimov dataset which we use as the observed dataset for the following studies.

Figure~\ref{fig:Limit} and Figure~\ref{fig:CLsb} show the results of one-sided and two-sided scanning using the gaussian fitting for several integrated luminosities. The left plot in Figure~\ref{fig:Limit} shows a median expected local p$_{0}$ under assumption of $H_{0}$ ($|V_{ts}|^2=0$) versus the true signal strength $\mu$. For $\mu = 1$, corresponding to the current SM expectation, the significance of rejecting $H_0$ is expected to be more than 5$\sigma$ when the integrated luminosity is 2000 \ifb and greater than 6$\sigma$ for 3000~\ifb, HL-LHC designed integrated luminosity. 
The right plot in Figure~\ref{fig:Limit} shows the median expected upper limit on $\mu$ under the assumption of $H_0$ at the 95\% confidence level (CL) for each luminosity as the dashed line and the $\pm 1(2) \sigma$ range as the green (yellow) band. Figure~\ref{fig:CLsb} shows the expected two-sided p-value distribution for the signal strength $\mu$ under the assumption of $\mu=1$.

Table~\ref{tab:upperlimit} summarizes several results shown above: the expected significance to exclude $\absvts=0$, the median 95 \% CL upper limit if $\mu=0$ and the 95\% CL for each luminosity.
Under the assumption of $\mu=1$, the expected significance from the hypothesis test calculation shows a 1.36$\sigma$ significance to reject the background-only hypothesis for the integrated luminosity of the Run2.
The value becomes 2.00$\sigma$ when the integrated luminosity is 300~\ifb, as is expected to be collected during Run 3 of the LHC, and 6.34$\sigma$ for the full HL-LHC integrated luminosity of 3000~\ifb.
Conversely, if the decay is suppressed, and $\mu=0$, then Run 3 of the LHC will be able to exclude $\mu=1$ at the 95\% CL level.

\begin{table*}[tb]
\centering
\begin{tabular}{|c|c|c|c|}
\hline
Integrated luminosity (\ifb) & \vtop{\hbox{\strut Expected significance ($\sigma$)} \hbox{\strut for $\absvts$ = 0 exclusion}} & \vtop{\hbox{\strut Expected 95\% CL}\hbox{\strut median upper limit ($\mu$)}} & \vtop{\hbox{\strut Expected 95\% CL$_{s+b}$} \hbox{\strut median interval ($\mu$)}} \\
\hline 
137.6 & 1.36 & $<$ 1.22 & [0.000, 2.30] \\
300 & 2.00 & $<$ 0.822 & [0.0210, 1.98] \\
600 & 2.83 & $<$ 0.582 & [0.307, 1.70] \\
1200 & 4.01 & $<$ 0.411 & [0.509, 1.49] \\
2000 & 5.17 & $<$ 0.319 & [0.619, 1.38] \\
3000 & 6.34 & $<$ 0.262 & [0.689, 1.31] \\
\hline
\end{tabular}
\caption{\label{tab:upperlimit} The expected significance to exclude $|V_{ts}|$ = 0, the expected 95\% CL median upper limit under the assumption of $|V_{ts}|=0$ and the expected 95\% CL$_{s+b}$ median confidence interval on $\mu = \frac{|V_{ts}^{obs}|^2}{|V_{ts}^{CKM}|^2}$ for several integrated luminosities. }
\end{table*}

\section{Conclusion}

We have studied the feasibility of a direct measurement of $|V_{ts}|$ from the dileptonic \ttbar production process, using hadronization by \pythia\ and \delphes\ detector simulation to produce a more realistic expectation of the results of the LHC experiments.
With an integrated luminosity of 3000~\ifb, which is expected to be achieved at the HL-LHC period, $\absvts$ = 0 can be excluded above the 6$\sigma$ significance level.

\section{Acknowledgments}

This article was supported by the computing resources of the GDSC at the Korea Institute of Science and Technology Information. 
W.J. is supported by the National Research Foundation of Korea (NRF) grant funded by the Ministry of Science and ICT (MSIT) (2018R1C1B6005826). 
J.L. is supported by the NRF grant funded by the MSIT (2019R1C1C1009200).
I.W. is supported by the Brain Pool Program through the NRF funded by the MSIT (2017H1D3A1A01052807).
I.P. is supported by the Basic Science Research Program through the NRF funded by the Ministry of Education (2018R1A6A1A06024977).

\bibliography{mybibfile}

\end{document}